\documentclass[doublecol]{epl2}

\title{The use of $\mu$-Bose gas model for effective modeling of dark matter}  
\shorttitle{The use of $\mu$-Bose gas model for modeling dark matter} 

\author{A.M. Gavrilik\footnote{E-mail: omgavr@bitp.kiev.ua}, I.I. Kachurik,
M.V. Khelashvili \and A.V. Nazarenko\footnote{E-mail:
nazarenko@bitp.kiev.ua}} \shortauthor{A.M. Gavrilik \etal}

\institute{
  \inst{} Bogolyubov Institute for Theoretical Physics - 14-b, Metrolohichna str.
 Kiev, 03143, Ukraine
}


 \abstract{
 For the recently introduced $\mu$-deformed analog of Bose gas model ($\mu$-Bose gas model), its
 thermodynamical aspects e.g. total number of particles and the partition function are certain
 functions of the parameter $\mu$. This basic $\mu$-dependence of thermodynamics of the
 $\mu$-Bose gas arises through the so-called $\mu$-calculus, an alternative to the known
 $q$-calculus (Jackson derivative, etc.), so we include main elements of $\mu$-calculus. Likewise,
 virial expansion of EOS and virial coefficients, the internal energy, specific heat and the
 entropy of $\mu$-Bose gas show $\mu$-dependence. Herein, we study  thermodynamical geometry of
 $\mu$-Bose gas model and find the singular behavior of (scalar) curvature, signaling for
 Bose-like  condensation. The critical temperature of condensation $T^{(\mu)}_c$ depending on
 $\mu$ is given and compared with the usual $T_c$, and with known $T_c^{(p,q)}$ of $p,q$-Bose
 gas model. Using the results on $\mu$-thermodynamics we argue that the condensate of $\mu$-Bose
 gas, like the earlier proposed infinite statistics system of particles, can serve for effective
 modeling of dark matter.}

\begin{document}

\maketitle

\section{Introduction}

  The works dealing with deformed Bose gas models are rather numerous,
 see e.g. \cite{Martin91,Manko,Chaichian,Monteiro,ShuChen02,
AdGavr,GavrSigma,AlginFibOsc,ScarfoneSwamy09,GR-intercepts,GM-correl,GR-12}.
 As usual, the deformations are based on respective deformed oscillator
 models such as $q$-oscillators~\cite{ArikCoon76,Biedenharn}
   and the 2-parameter $p,q$-deformed (or Fibonacci) oscillators
     \cite{Chakrabarti91}.
  A plenty of nonstandard one-parameter deformed oscillators (DOs) exist~\cite{Plethora08},
  along with polynomially deformed ones~\cite{Polynomially}.
  Among so-called quasi-Fibonacci oscillators~\cite{Kachurik}
 we find the $\mu$-deformed oscillator~\cite{Jann}.
  The DO models often possess unusual properties, e.g. various energy level
  degeneracies, nontrivial recurrent relations for energy spectra etc.
 Nontrivial features of DOs motivate their application in diverse fields of
quantum physics.

 Physical meaning of deformation and its parameter(s) depend on specific
 application of the model of deformed bosons to a physical system.
 When the model of ideal gas of {\it deformed} bosons is applied to computing
 the intercepts of the momentum correlation functions~\cite{AdGavr,GavrSigma,GR-intercepts},
 one effectively takes into account the non-zero proper volume of
 particles~\cite{av95}, or their internal structure or compositeness~\cite{GKM2011,GM2012}.
 The experimental data on the two-pion correlation-function intercepts shows~\cite{Abelev}
 non-Bose type behavior of pions, and the efficiency of using deformed analogs
 of Bose gas model in that context has been demonstrated~\cite{GavrSigma,AGP,GM_NP}.
  There exists a successful application of $q$-Bose gas setup to overcome the
  difficulty with unstable phonon spectrum, confirmed~\cite{Monteiro} by
  experimental measurements of phonon lifetime. 

 In the $\mu$-Bose gas model and other deformed analogs of Bose gas model, quantum
 statistical interaction gets modified~\cite{GR-12,AlginSenay}.    
 The deformation may also absorb~\cite{ScarfoneSwamy09}
 an interaction present in the initially non-deformed system.

 The $\mu$-Bose gas model was proposed in~\cite{GR-intercepts,GM-correl}
 wherein the (intercepts of) correlation functions of 2nd,
 and higher order were derived.
 The study of $\mu$-Bose gas thermodynamics started in~\cite{RGK}
 used the special so-called $\mu$-calculus.
  That allowed to explore important quantities and aspects.

 In this paper, we wish to explore whether $\mu$-Bose gas model
 is able to model basic features of dark matter, and make some first steps.
 Among the approaches to model dark matter, those
based on the idea of Bose-Einstein condensate (BEC) are very
 numerous, see e.g.\cite{Sin,Boehmer,Harko,Arbey,Hu,Kain}, review~\cite{review}
 and references therein: these possess plausible
features of cold dark matter, in some respects show advantages,
though also may confront with their own difficulties, say the
problem of gravitational collapse\cite{Guzman}.

In view of unknown precise nature of dark matter constituents,
rather exotic candidates were proposed, e.g. axionic~\cite{axion} or
even stringy ones~\cite{stringy}. In some papers, the authors
exploited different models of nonstandard thermostatistics aimed at
a modeling of unusual physical objects in quantum cosmology,
 e.g. in the physics of dark matter~\cite{Infinite,Dil}
 or black holes~\cite{Strominger,Ng,Zare}.
 But, we think it is worth to examine other potentially interesting
 models of nonstandard thermostatistics as candidates for modeling,
 at least effectively, main properties of dark matter -- in order to
 choose most appropriate one.

 About the plan of our paper.
 In Sec.~2 we give main facts concerning the $\mu$-deformed
 Bose gas model basing on $\mu$-calculus.
 Then in Sec.~3 the thermodynamic quantities are explored.
 The expression for the total number of particles allows to
 derive the partition function (both quantities carry explicit $\mu$-dependence).
  In our study we mainly explore the regime of low temperature.
  In Sec.~4, we use the geometric approach to thermodynamics
  (see e.g.~\cite{Ruppeiner,Janyszek,Ubriaco}) and explore the
  possibility of Bose like condensation in the $\mu$-Bose gas model.
  The critical temperature of condensation and its dependence
  on the (deformation) parameter $\mu$ are studied.
   Other aspects and thermodynamical functions are considered in the next section.
   Discussion of possible application to dark matter as well as concluding remarks
   are given in the final section of the paper.

\section{Deformed analogs of Bose gas model}
Like in other works on deformed oscillators, see e.g.
\cite{Monteiro,ScarfoneSwamy09,AlginFibOsc} we deal in fact with the
(system of) deformed bosons.
 The virtues of such deformation is its ability to provide effective account
 of interaction between particles, their non-zero volume, their
 inner (composite) structure etc.

 Deformed Bose gas model termed $\mu$-Bose gas model
 associated with $\mu$-deformed oscillators        \cite{Jann}
 was developed                       in~\cite{GR-intercepts, GM-correl}.
 Therein, and in this paper, the thermal
average of the operator $\mathcal{O}$  is determined by the familiar
formula
\begin{equation}\label{eq.2}
\langle \mathcal{O} \rangle=\frac{Tr(\mathcal{O}e^{-\beta H})}{Z},
\end{equation}
$Z$ being the grand canonical partition function.
 Its logarithm is
\begin{equation}\label{eq.3}
\ln Z=-\sum_i\ln(1-ze^{-\beta\varepsilon_i})
\end{equation}
with the fugacity $z=e^{\beta\widetilde{\mu}}$.
 The familiar formula
\begin{equation}\label{eq.4}
N=z\frac{d}{dz}\ln Z
\end{equation}
  for the number of particles will be modified, see below.

 To study deformed Bose gas model, namely the $\mu$-Bose
gas as the model describing the system of deformed bosons, we use
 the Hamiltonian ($\widetilde{\mu}$ is chemical potential)
  \begin{equation}\label{eq.1}
H=\sum_i(\varepsilon_i-\widetilde{\mu})N_i
\end{equation}
Here $\varepsilon_i$ denotes kinetic energy of particle in the state
"$i$",  $N_i$ the particle number (occupation number) operator
corresponding to state "$i$".
 To develop thermodynamics in the $\mu$-deformed model,
  we need the so-called $\mu$-calculus.

\subsection{Elements of $\mu$-calculus}
  The familiar path of deriving thermodynamical functions and relations
  implies usage of standard derivative $d/dx$.  
  Since we wish to develop the $\mu$-analog of the Bose gas model,
  we extend (deform) the very notion of derivative.
   The so-called $\mu$-derivative, introduced     in~\cite{RGK},
  differs from the known Jackson or          $q$-derivative~\cite{Kac}
  and its $p,q$-extension (used in~\cite{GR-12}).
  The easiest way to introduce the $\mu$-extension
  is by the rule
    \begin{equation}\label{eq.10}
\mathcal{D}^{(\mu)}_x x^n\!=\![n]_{\mu}x^{n\!-\!1},  \quad
 [n]_{\mu}\!\equiv\!\frac{n}{1+\mu n} \quad \mbox{($\mu$-bracket)}\,
 \end{equation}
 so that the $\mu$-derivative involves the $\mu$-bracket
 from the work on $\mu$-oscillator\cite{Jann}.
  If $\mu\!\rightarrow\! 0$, $[n]_{\mu}\to n$, and the $\mu$-extension
$\mathcal{D}^{(\mu)}_x$ reduces to usual derivative $d/dx$.

 Knowledge of the action of $\mu$-derivative on monomials $x^m$
 is enough for the goals of this work, while
  the general rule for action of
  $\mu$-derivative on general function $f(x)$ is
    \begin{equation}\label{eq.11}
\mathcal{D}^{(\mu)}_xf(x)=\int^1_0dtf'_x(t^{\mu}x),  \qquad
f'_x(t^{\mu}x)=\frac{d f(t^{\mu}x)}{d x}.
\end{equation}
 Clearly, formula (\ref{eq.10}) stems from this general definition.

 For $k$th power of $\mu$-derivative acting on $x^n$ we have
\begin{equation}\label{eq.12}
(\mathcal{D}^{(\mu)}_x)^kx^n=\frac{[n]_{\mu}!}{[n-k]_{\mu}!}x^{n-k},
\qquad [n]_{\mu}!\equiv\frac{n!}{(n;\mu)}
\end{equation}
where $(n;\mu)\equiv(1+\mu)(1+2\mu)...(1+n\mu)$.

 The inverse $\bigl({\mathcal{D}^{(\mu)}_x}\bigr)^{-1}$
 of the $\mu$-derivative $D^{(\mu)}_x$ in (\ref{eq.10}) and (\ref{eq.11})
 can as well be defined.

 Let us note that for the $\mu$-derivative (\ref{eq.11}) there exist its $q,{\mu}$- or
$(p,q;\mu)$-deformed extensions: instead of $(d/dx)f(t^{\mu}x)$ in
(\ref{eq.11}) one takes $\mathcal{D}^{q}_xf(t^{\mu}x)$ or
$\mathcal{D}^{(p,q)}_xf(t^{\mu}x)$. The two extensions correspond to
the $(q;\mu)$- or $(p,q;\mu)$-deformed quasi-Fibonacci oscillators
in~\cite{Kachurik}.

 So, to develop the $\mu$-Bose gas thermodynamics we apply, where necessary,
 the $\mu$-derivative $D^{(\mu)}_z$ instead of usual $d/dz$.
 Due to $\mu$-derivative, the parameter $\mu$ of deformation
 enters the treatment, and the system becomes $\mu$-deformed.
    For small $\mu \ll 1$, the usual and the deformed
 derivatives of a function have similar behavior, that
  is easily seen by acting with $\mu$-derivative and usual one
 on the monomial, logarithmic, exponential function, etc.
  Such property of $\mu$-derivative justifies, at least partly,
  its use in developing thermodynamics of $\mu$-Bose gas.

 Appearance of $\mu$-bracket $[n]_{\mu}$ and $\mu$-factorial
$[n]_{\mu}!$, see (\ref{eq.10}), (\ref{eq.12}) generates
$\mu$-deformed analogs~\cite{RGK} of elementary functions:
 $\mu$-exponential $exp_{\mu}(x)$, $\mu$-logarithm $\ln_{\mu}(x)$
 (involving $\mu$-numbers $[n]_{\mu}$ and $\mu$-factorial
$[n]_{\mu}!=[n]_{\mu}[n-1]_{\mu}...[2]_{\mu}[1]_{\mu}$).
 New special functions e.g., $\mu$-analog of polylogarithms,
 do also appear, see~\cite{RGK} and below.

\subsection{Acting on product of functions by $D^{(\mu)}_x$
 ($\mu$-Leibnitz rule)}
 Let the $\mu$-derivative act on the product
$f(x)\cdot g(x)$.  
 From definition $(10)$, taking the monomials $f(x)=x^n$ and
$g(x)=x^m$ we infer                   
\begin{equation}        \label{eq.13}
  \hspace{-2mm}  D^{(\mu)}_x \left( x^n x^m\right)\!=\! D^{(\mu)}_x \left( x^m
x^n\right)\!=\! \frac{n\!+\!m}{1+\mu(n\!+\!m)}\, x^{n+m-1} .
\end{equation}
 Formula for $D^{(\mu)}_x$ operating on general product
 $f(x)\cdot g(x)$ does also exist, but we will not need it in this work.

\section{Thermodynamics of $\mu$-Bose gas model}

Now we study thermodynamics of $\mu$-Bose gas model using the
 $\mu$-calculus.
 Consider the gas of non-relativistic particles and focus mainly
 on the regime of low temperatures. 

\subsection{Total number of particles}
 The usual relation for total number of Bose gas particles is
\begin{equation}\label{eq.14}
N=z\frac{d}{dz}\ln Z.
\end{equation}
 For $\mu$-Bose gas thermodynamics, the
total number of particles $N\equiv N^{(\mu)}$ is defined as
\begin{equation}\label{eq.15}
N^{(\mu)}=z\mathcal{D}^{(\mu)}_z\ln Z
=-z\mathcal{D}^{(\mu)}_z\sum_i\ln(1-ze^{-\beta\varepsilon_i})
\end{equation}
where the $\mu$-derivative $\mathcal{D}^{(\mu)}$ from (\ref{eq.10})
is used.
 For $\mu\geq 0$, we apply it to the ($\log$ of) partition
function in (\ref{eq.3}) to get
\begin{equation}\label{eq.16}
N^{(\mu)}\!=\!z\sum_i\!\sum_{n=1}^{\infty}\frac{e^{-\beta\varepsilon_in}}{n}
   [n]_{\mu}z^{n-1}\!=\!
 \sum_i\!\sum_{n=1}^{\infty}\frac{[n]_{\mu}}{n}(e^{-\beta\varepsilon_i})^nz^n\, .
\end{equation}
We require $0\leq|ze^{-\beta\varepsilon_i}|<1$ in (\ref{eq.16}).
 As we deal with non-relativistic particles of mass $m$,
 the energy $\varepsilon_i$ is taken as
\begin{equation}\label{eq.17}
\varepsilon_i=\frac{\overrightarrow{p}_i\overrightarrow{p}_i}{2m}
=\frac{|p|^2}{2m}=\frac{p_i^2}{2m}
\end{equation}
with the 3-momentum $\overrightarrow{p}_i$ of particle in $i$-th
state.

 At $z\rightarrow 1$ the summand in (\ref{eq.16}) diverges when $p_i=0, i=0$.
 So  we assume that the $i=0$ ground state admits
  macroscopically large occupation number.
  Likewise, even for $z\neq 1$ we as well separate
  the term with $p_i=0$ from the remaining sum:
\begin{equation}\label{eq.18}
N^{(\mu)}={\sum_i} ' \sum_{n=1}^{\infty}\frac{[n]_{\mu}}{n}(e^{-\beta\varepsilon_i})^nz^n +
\sum_{n=1}^{\infty}\frac{[n]_{\mu}}{n}z^n.
\end{equation}
The sum symbol ${\sum_i}'$ in (\ref{eq.18}) means that the $i=0$
term is dropped from the sum.
  For large volume $V$ and large $N$ the spectrum of single-particle
  states is almost continuous so we replace the sum in (\ref{eq.16}) by integral:
\begin{equation}\label{eq.19}
\sum_i\rightarrow \frac{V}{(2\pi \hbar)^3}\int d^3k.
\end{equation}
Thus, we isolate the ground state and include the contribution from
all other states in the integral.
 To compute the total number of particles we integrate
 over 3-momenta in spherical coordinates:
 \begin{equation}\label{eq.20}
  N^{(\mu)}\!=\!\frac{4\pi V}{(2\pi\hbar^2)^3}
  \!\sum_{n=1}^{\infty}\!\frac{[n]_{\mu}z^n}{n}\!\!\int^{\infty}_0
  \!\!\!\!p^2e^{-\frac{\beta p^2}{2m}}dp +\!\sum_{n=1}^{\infty}\!\frac{[n]_{\mu}z^n}{n}.
 \end{equation}
 In the integral, the lower limit can still be taken as zero:
 indeed, the ground state, $p_0$, does not contribute to the integral.
  Integration by parts leads us to the $\mu$-deformed total number
  of particles:
\begin{equation}\label{eq.21}
N^{(\mu)}=\frac{V}{\lambda^3}\sum_{n=1}^{\infty}\frac{[n]_{\mu}}{n^{5/2}}z^n+
N_0^{(\mu)}, \quad N_0^{(\mu)}\equiv
\sum_{n=1}^{\infty}\frac{[n]_{\mu}}{n}z^n ,
\end{equation}
where $\lambda=\sqrt{\frac{2\pi \hbar^2}{mkT}}$ is the thermal wavelength.
 This result can be presented through the $\mu$-analog of Bose function:
\begin{equation}\label{eq.22}
N^{(\mu)}=\frac{V}{\lambda^3}g_{3/2}^{(\mu)}(z)+g_0^{(\mu)}(z),
\qquad  g_0^{(\mu)}(z)=N_0^{(\mu)} .
\end{equation}
 Here $g_0^{(\mu)}(z)$ and $g_{3/2}^{(\mu)}(z)$ are
 the $\mu$-polylogarithm , i.e. $\mu$-analog of the usual polylogarithm
  $g_l(z)=\sum_{n=1}^{\infty}z^n/n^l$:
\begin{equation}\label{eq.23}
g_{l}^{(\mu)}(z)=\sum_{n=1}^{\infty}\frac{[n]_{\mu}}{n^{l+1}}z^n.
\end{equation}
 The limit $\mu\rightarrow 0$ recovers usual $g_l(z)$ function (polylogarithm).

 For positive real $\mu$ the convergence properties are not spoiled and,
 like for the standard $g$-function $g_l(z)$, there should be $|z|<1$.
 Note that the restriction $z<1/q$ does
 appear for the version of $q$-Bose gas studied in Ref.~\cite{ShuChen02}.

For the needs of subsequent analysis, let us rewrite the expression
in (\ref{eq.22}) for total number of particles as
\begin{equation}\label{eq.24}
\frac{1}{v}=\frac{1}{\lambda^3}g^{(\mu)}_{3/2}+\frac{N_0^{(\mu)}}{V},
\qquad v\equiv\frac{V}{N^{(\mu)}}.
\end{equation}

\subsection{Deformed grand partition function}
 In $\mu$-Bose gas model, we use the relations between thermodynamical functions
 similar to those of usual Bose gas thermodynamics,
 but in our case all the thermodynamical functions
 including the partition function 
 become $\mu$-dependent.

 To obtain deformed partition function $\ln Z^{(\mu)}$ take
\begin{equation}\label{eq.25}
N^{(\mu)}=z\frac{d}{dz}\ln Z^{(\mu)}
\end{equation}
and invert it:
 \begin{equation}\label{eq.26}
\ln Z^{(\mu)}=\Bigl(z\frac{d}{dz}\Bigr)^{-1}N^{(\mu)}.
\end{equation}
To apply $\bigl(z\frac{d}{dz}\bigr)^{-1}$,
   we use the following property for a function $f(z\frac{d}{dz})$
   admitting power series expansion:
\begin{equation}\label{eq.27}
f\Bigl(z\frac{d}{dz}\Bigr)z^k=f(k)z^k.
\end{equation}
 Then, from (\ref{eq.26}), (\ref{eq.27}) and
(\ref{eq.21}) we infer
$$
\ln Z^{(\mu)}=\Bigl(z\frac{d}{dz}\Bigr)^{-1}
 \left(\frac{V}{\lambda^3}\sum_{n=1}^{\infty}\frac{[n]_{\mu}}{n^{5/2}}z^n+
\sum_{n=1}^{\infty}\frac{[n]_{\mu}}{n}z^n\right)=
$$
\begin{equation}\label{eq.28}
=\frac{V}{\lambda^3}
 \sum_{n=1}^{\infty}\frac{[n]_{\mu}}{n^{5/2}}(n)^{-1}z^n
 +\sum_{n=1}^{\infty}\frac{[n]_{\mu}}{n}(n)^{-1}z^n
\end{equation}
or, in a more compact form,
   \begin{equation}                 \label{eq.29}
  \hspace{-35mm}    \ln
  Z^{(\mu)}=\frac{V}{\lambda^3}g^{(\mu)}_{5/2}+g^{(\mu)}_1\, ,
       \vspace{-2mm}
\end{equation}
   \begin{equation}                 \label{eq.30}
    \vspace{-1mm}
  Z^{(\mu)}(z,T,V)=\exp\biggl(\frac{V}{\lambda^3}g^{(\mu)}_{5/2}(z)+g^{(\mu)}_1(z)\biggr).
\end{equation}
 Formulas (\ref{eq.28})-(\ref{eq.30}) provide the $\mu$-deformed
 partition function and play basic role: using (\ref{eq.30})
 we  can derive other thermodynamical functions and relations.

\section{Geometric approach to $\mu$-Bose gas model}

Let us study the thermodynamics of $\mu$-Bose gas using
thermodynamical geometry in the space with two   
parameters $\beta, \gamma$, where $\gamma=-\beta\tilde{\mu}$ and
$\tilde{\mu}$ is chemical potential.
 The components of the metric in the Fisher-Rao representation are defined as
\begin{equation}\label{metbb}
G_{\beta\beta} = \frac{\partial^2\ln Z^{(\mu)}}{\partial \beta^2} =
-\left(\frac{\partial U}{\partial \beta}\right)_{\gamma} \, ,
\end{equation}
\begin{equation}\label{metbg}
G_{\beta\gamma} = \frac{\partial^2\ln Z^{(\mu)}}{\partial\gamma \,
\partial\beta} = -\left(\frac{\partial N}{\partial
\beta}\right)_{\gamma},
\end{equation}
\begin{equation}\label{metgg}
  G_{\gamma\gamma} = \frac{\partial^2\ln Z^{(\mu)}}{\partial\gamma^2} =
-\left(\frac{\partial N}{\partial \gamma}\right)_{\beta}
\end{equation}
(note that $G_{\gamma\beta}=G_{\beta\gamma}$).
 Using these and eq.~(\ref{eq.29}) we calculate the expressions
 for metric components and obtain
\begin{equation}
G_{\beta\beta} =
 \frac{15}{4}\frac{V}{\lambda^3\beta^2}\,g^{(\mu)}_{\frac{5}{2}}(z)\, ,
\end{equation}
\begin{equation}
G_{\beta\gamma}  =
\frac{3}{2}\frac{V}{\lambda^3\beta}\,g^{(\mu)}_{\frac{3}{2}}(z) \, ,
\end{equation}
\begin{equation}
G_{\gamma\gamma} = \frac{V}{\lambda^3}\,g^{(\mu)}_{\frac{1}{2}}(z) +
g^{(\mu)}_{-1}(z)  \, .
\end{equation}
From (29)-(31) the determinant of the metric results:
$$
det|G_{ij}| \equiv g =
\frac{3}{4}\frac{V}{\lambda^3\beta^2}\left(5g^{(\mu)}_{\frac{5}{2}}(z)
g^{(\mu)}_{-1}(z) + \right. \hspace{1.2cm}
$$
\begin{equation}
\hspace{10mm} +\left. \frac{V}{\lambda^3}\left( 5
g^{(\mu)}_{\frac{5}{2}}(z)\,g^{(\mu)}_{\frac{1}{2}}(z) -
3g^{(\mu)}_{\frac{3}{2}}(z)\,g^{(\mu)}_{\frac{3}{2}}(z)
\right)\right)   \, .
\end{equation}
Components of inverse metric are given as ($\beta\leftrightarrow 1,
\, \gamma\leftrightarrow 2$)
\begin{equation}
G^{11} = \frac{G_{22}}{g} \, , \qquad G^{12} = - \frac{G_{12}}{g} \,
, \qquad G^{22} =  \frac{G_{11}}{g}  \, .
\end{equation}
 Since the metric components are expressed through derivatives of partition
 function, see (\ref{metbb})-(\ref{metgg})\,),
  the formulas for Christoffel symbols and the Riemann tensor
  are found as
   \begin{equation}
 \hspace{-26mm}
\Gamma^{(\mu)}_{\lambda\sigma\nu} = \frac{1}{2}\bigl(\ln
Z^{(\mu)}\bigr)_{, \lambda\sigma\nu} \, .
\end{equation}
\begin{equation}
R_{\lambda\sigma\nu\rho} \equiv
g^{\kappa\tau}\left(\Gamma_{\kappa\lambda\rho}\Gamma_{\tau\sigma\nu}
- \Gamma_{\kappa\lambda\nu}\Gamma_{\tau\sigma\rho} \right)  \, .
\end{equation}
Calculation of the Christoffel symbols yields
$$
\Gamma_{\beta\beta\beta} =
 -\frac{105}{16}\frac{V}{\lambda^3\beta^3}\,g^{(\mu)}_{\frac{5}{2}}(z)
  \, ,
$$
$$
\Gamma_{\beta\beta\gamma} =
-\frac{15}{8}\frac{V}{\lambda^3\beta^2}\,g^{(\mu)}_{\frac{3}{2}}(z)
 \, ,
$$
$$
\hspace{-4mm} \Gamma_{\beta\gamma\gamma} =
-\frac{3}{4}\frac{V}{\lambda^3\beta}\,g^{(\mu)}_{\frac{1}{2}}(z)
 \, ,
$$
$$
\Gamma_{\gamma\beta\beta} =
-\frac{15}{8}\frac{V}{\lambda^3\beta^2}\,g^{(\mu)}_{\frac{3}{2}}(z)
 \, ,
$$
$$
\hspace{-18mm}
 \Gamma_{\gamma\beta\gamma} =
-\frac{3}{4}\frac{V}{\lambda^3\beta}\,g^{(\mu)}_{\frac{1}{2}}(z)
 \, ,
$$
$$
\Gamma_{\gamma\gamma\gamma} =
-\frac{1}{2}\left(\frac{V}{\lambda^3}\,g^{(\mu)}_{-\frac{1}{2}}(z) +
g^{(\mu)}_{-2}(z) \right) \, .
$$
As known, in 2-dimensional space the scalar curvature $R$ is
determined by one component of Riemann tensor, i.e.
 \begin{equation}        \label{Rdef}
R = \frac{2R_{1212}}{g}    \, .
\end{equation}
Denoting $ g_l^{(\mu)}(z)\equiv g_l^\mu$, for the component
$R_{1212}$ we obtain:
$$
R_{\beta\gamma\beta\gamma}\!=
\frac{45}{64}\left(\frac{V}{\lambda^3\beta^2}\right)^2\frac{1}{g}
\left[ 5g^{\mu}_{\frac{3}{2}} g^{\mu}_{\frac{3}{2}} g^{\mu}_{-1} -
7g^{\mu}_{\frac{5}{2}} g^{\mu}_{\frac{1}{2}} g^{\mu}_{-1} +\right.
$$
$$ \ \
+\, 2g^{\mu}_{\frac{3}{2}} g^{\mu}_{\frac{5}{2}} g^{\mu}_{-2} +
\frac{V}{\lambda^3}\left( 2g^{\mu}_{\frac{3}{2}}
g^{\mu}_{\frac{5}{2}} g^{\mu}_{-\frac{1}{2}} -4g^{\mu}_{\frac{5}{2}}
g^{\mu}_{\frac{1}{2}} g^{\mu}_{\frac{1}{2}} + \right.
$$
\begin{equation}
\left.\left. \hspace{-42mm} +\, 2g^{\mu}_{\frac{3}{2}}
g^{\mu}_{\frac{3}{2}} g^{\mu}_{\frac{1}{2}} \right) \right] .
\end{equation}
Substitution of this in eq.~(\ref{Rdef}) yields our final result:
$$  \hspace{-5mm}
R = \frac52 \, 
\biggl( 5g^{\mu}_{\frac{3}{2}}
g^{\mu}_{\frac{3}{2}} g^{\mu}_{-1} -7g^{\mu}_{\frac{5}{2}}
g^{\mu}_{\frac{1}{2}} g^{\mu}_{-1} + +2g^{\mu}_{\frac{3}{2}}
g^{\mu}_{\frac{5}{2}} g^{\mu}_{-2} + \biggr.
$$
$$
\hspace{-4mm} \biggl.+\, \frac{V}{\lambda^3}\left(
2g^{\mu}_{\frac{3}{2}} g^{\mu}_{\frac{5}{2}} g^{\mu}_{-\frac{1}{2}}
-4g^{\mu}_{\frac{5}{2}} g^{\mu}_{\frac{1}{2}} g^{\mu}_{\frac{1}{2}}
+2g^{\mu}_{\frac{3}{2}} g^{\mu}_{\frac{3}{2}} g^{\mu}_{\frac{1}{2}}
\right) \biggr)\times
$$
\begin{equation} \hspace{-9mm}
\times\left(5g^{\mu}_{\frac{5}{2}} g^{\mu}_{-1} +
\frac{V}{\lambda^3}\left( 5
g^{\mu}_{\frac{5}{2}}\,g^{\mu}_{\frac{1}{2}} -
3g^{\mu}_{\frac{3}{2}}\,g^{\mu}_{\frac{3}{2}} \right)\right)^{-2} .
\end{equation}
Curvature in the thermodynamical parameters space is a useful tool
to study thermodynamical properties of the system: it becomes
singular at the phase transition points.
 The $\mu$-polylogarithm $g^{(\mu)}_l(z)$ involved in curvature $R$
 is singular at
 $z\to 1$, for $l \le 1$ in the case $\mu = 0$ and for $l \le 0$ when
$\mu \ne 0$. Thermodynamical curvature $R(z)$ in isothermal process
has characteristic properties as a function of fugacity. In the case
$V/\lambda^3 \ll 1 $ the curvature $R(z)$ has no singularities as
seen in Fig.~1.

 In the case $V/\lambda^3 \gg 1 $ the dependence looks as shown in Fig.~2.
 If $V/\lambda^3$ is sufficiently large to neglect the terms which do not
 contain this factor, the curvature $R(z)$ is singular at $z\to 1$.
   We conclude that, in the latter situation, the system undergoes phase
   transition, and hence Bose-like condensation takes place. That is,
  the $\mu$-Bose gas model satisfies basic necessary property.

\begin{figure}[h]
\onefigure[width=0.9 \linewidth]{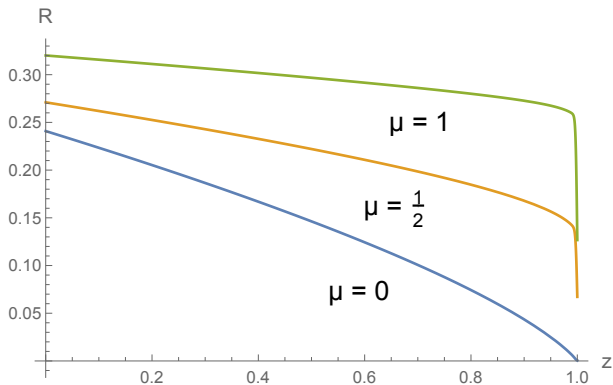}
 \caption{Scalar curvature R(z) in the case $V/\lambda^3 \ll 1 $, for
 different values of deformation parameter $\mu = 0, \frac{1}{2}, 1 $,
 in isothermal process $\beta = const$.}
 \label{fig.1}
\end{figure}

\begin{figure}[h]
\onefigure[width=0.9\linewidth]{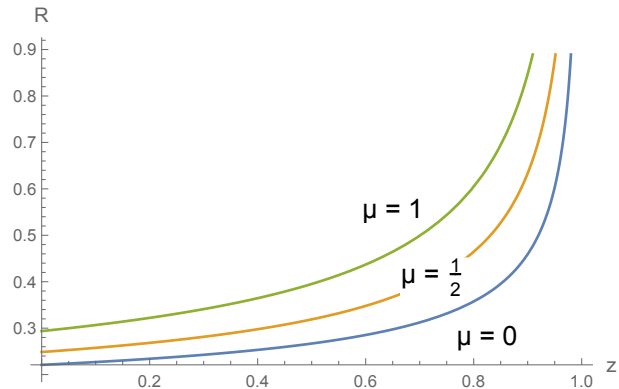}
 \caption{Scalar curvature R(z) in the case $V/\lambda^3 \gg 1 $, for
 different values of deformation parameter $\mu = 0, \frac{1}{2}, 1 $,
 in isothermal process $\beta = const$.}
 \label{fig.2}
\end{figure}

 \vspace{1mm}
\section{Critical temperature of condensation}
In the regime of low temperature and high density one can obtain,
like  for $p,q$-Bose gas \cite{GR-12}, the critical temperature
$T^{(\mu)}_c$ of condensation in the considered $\mu$-deformed
 Bose gas model \cite{RGK}. We start with eq.~(\ref{eq.24})
and rewrite it as
\begin{equation}\label{eq.36}
\frac{N_0^{(\mu)}}{V}=\frac{\lambda^3}{v}-g^{(\mu)}_{3/2}(z).
\end{equation}
  The critical temperature $T_c^{(\mu)}$ of $\mu$-Bose gas is
determined by the equation $\lambda^3/{v}=g^{(\mu)}_{3/2}(1)$, that
gives \cite{RGK}:
\begin{equation}\label{eq.37}
T_c^{(\mu)}=\frac{2\pi\hbar^2/mk}{\bigl(vg^{(\mu)}_{3/2}(1)\bigr)^{2/3}}.
\end{equation}
Using the latter we infer the ratio of critical temperature
$T_c^{(\mu)}$ to the critical temperature $T_c$ of usual Bose
gas~\cite{Pathria}:
\begin{equation}\label{eq.38}
\frac{T_c^{(\mu)}}{T_c}=\Biggl(\frac{2.61}{g^{(\mu)}_{3/2}(1)}\Biggr)^{2/3}.
\end{equation}
  We see that the ratio $T_c^{(\mu)}/{T_c}$, like in the case of
$p,q$-Bose gas model \cite{GR-12}, has an important feature: the
greater is the strength of deformation (given by $\mu$) the higher
is $T_c^{(\mu)}$ (say, for $\mu=0.06$ we have $T_c^{(\mu)}\simeq
1.22 \cdot T_c$).
 If $\mu\!\rightarrow\!0$ (no-deformation limit), the ratio $T_c^{(\mu)}/{T_c}=1$,
 i.e, the $\mu$-critical temperature tends to usual one,
$T_c^{(\mu)}\rightarrow T_c$ (a kind of consistency).
   The existence of condensate of $\mu$-bosons is {\it the first and crucial
condition} for possible use of $\mu$-Bose gas model in the
(effective) modeling of dark matter.

Let $T$ be in the interval $0\!<\!T\!<\!T^{(0)}_c\!\leq
T^{(\mu)}_c$.   In the $\mu$-deformed case we have
$\frac{U^{(\mu)}}{T}=\frac25 c^{(\mu)}_v= \frac35 S^{(\mu)}$ in
similarity with pure Bose case, i.e. $\frac{U}{T}=\frac25 c_v=
\frac35 S .$  From these we infer two useful relations:\
$\frac{U^{(\mu)}}{c^{(\mu)}_v}=\frac{U}{c_v}$
 and $\frac{U^{(\mu)}}{S^{(\mu)}}=\frac{U}{S}$.

\section{Possible application for modeling dark matter}
  In this paper, using the tools of thermodynamic geometry, we
verified the properties of $\mu$-Bose gas model needed for its
modeling the dark matter properties. The appearance of Bose-like
condensation as main property, is confirmed.

As emphasized in \cite{Harko}, the dark matter surrounding dwarf
galaxies is to be viewed as "{\it strongly coupled, dilute system of
particles}". In the case of $\mu$-Bose gas, we should stress that
the interaction between $\mu$-bosons (of pure quantum statistical
origin) is also attractive, and due to deformation can be even
stronger than that for pure bosons. To see that compare the two 2nd
virial coefficients: $\mu$-dependent one $V_2^{(\mu)}$~\cite{RGK},
and the standard $V_2^{(Bose)} = 2^{-5/2}$ (drop the sign "minus"
for the both):
$$
V_2^{(\mu)}-{V_2}^{Bose}= 2^{-7/2} \biggl(
\frac{[2]_{\mu}}{[1]^2_{\mu}} - 2 \biggr)
=2^{-5/2}\frac{\mu^2}{1+2\mu} > 0 .
$$
I.e., due to enhanced attraction (of quantum origin) we may say than
the $\mu$-bosons are "more bosonic" than usual bosons. This property
is good for providing {\it strongly coupled} system of
(quasi)particles.
We see that {\it at large} $\mu$ one has
$g^{(\mu)}_l(z)\to\mu^{-1}g^{(0)}_l(z)\equiv\mu^{-1}g_{l+1}(z)$,
where $g_l(z)$ is the polylogarithm function. Then, the internal
energy per particle would not depend on $\mu$ while the total one
does, because of scale factor.

However, the unlimited growth of the parameter $\mu$ and thus of the
strength of attraction, could lead to a collapse of the quantum
system under study. To prevent that, we can find some bound on the
values of $\mu$. Namely, the requirement to forbid negative pressure
can do the job. We take  (virial expansion of) the equation of
state~\cite{RGK} to second order i.e.
\begin{equation}
 \frac{P v}{k T}=1-\frac{\ \ \ [2]_{\mu}}{2^{7/2}[1]^2_{\mu}}\frac{\lambda^3}{v
 }\, .
\end{equation}
Imposing $P=0$ yields the relation for finding critical value
$\bar{\mu}$ of deformation strength: $\frac{2^{-7/2}
[2]_{\mu}}{[1]^2_{\mu}}\frac{\lambda^3}{v} = 1$.
 From that,
 \begin{equation}
   \bar{\mu} = \kappa -1 + \sqrt{(\kappa -1)\kappa},
    \hspace{10mm}
  \kappa
 \equiv 2^{5/2} \frac{v}{\lambda^3}\, ,
 \end{equation}
and we have the condition $\mu \leq \bar{\mu}$ to avoid the
collapse. Obviously, $\kappa\!=\!1$ yields $\bar{\mu}=0$. Then
$\mu=0$ and we recover pure Bose case.
 On the other hand, the bound taken say as $\bar{\mu}=1$ with
 $0 < \mu \le 1$ corresponds to the value $\kappa=\frac{4}{3}$.

 There exists a special "characteristic" value of ${\mu}$
for which the deformed entropy (see eq.~(41) and Fig.~6 in [27])
becomes: \ $ \frac{S^{(\mu_0)}\lambda^3}{V k_B} =1$ \ or just \  $
\frac{S^{(\mu_0)}\lambda^3}{V} = k_B$.
 We take this special value of $\mu$ as the bound $\bar{\mu}$ of our interval,
 so that $0<\mu<\bar{\mu}\!\equiv\!{\mu}_0\!=\!1.895$.
  At this deformation strength ${\mu}_0\simeq 1.895$, we obtain
  the relation
 \begin{equation}
  g_{3/2}(1) = 3.3535 \ g_{3/2}^{(\mu=\mu_0)}(1)
 \end{equation}
for usual polylogarithm and $\mu$-polylogarithm. From that, we find
that the critical volume-per-particle (divided by cube of thermal
wavelength) in the $\mu$-Bose gas is related with similar quantity
of the usual Bose gas by the formula:
$$ \Bigl(\frac{v^c}{\lambda^3}\Bigr)_{\mu = \mu_0}=
  3.3535\, \Bigl(\frac{v^c}{\lambda^3}\Bigr)_{Bose} \, . $$
 Then we can estimate the ratio in our model using
respective estimates from Bose-condensate case, see
e.g.~\cite{Harko}.

Using the BEC model of dark matter by Harko, based on the
Gross-Pitaevskii equation in the Thomas-Fermi approximation, we
deduce the characteristics of dark matter halos modified due to
$\mu$-deformed statistics. Introducing the dimensionless factor
  $$  
f=\sqrt{\frac{2\pi akT}{Gm^2}}\gg1
  $$  
where $a$ is the $s$-wave scattering length which is less than
thermal wavelength ($a<\lambda$), $G$ is the gravitational constant,
we find the total mass of halo and its radius:
\begin{equation}
M^{(\mu)}=\frac{\pi}{6}m g^{(\mu)}_{3/2}(1)f^3,\qquad
R=\frac{1}{2}\lambda f=\pi\sqrt{\frac{\hbar^2a}{Gm^3}}.
\end{equation}
Since $g^{(\mu)}_{3/2}(1)<g^{(0)}_{3/2}(1)$ at $\mu>0$, see also
eq.~(44), we may expect a {\it better agreement of our predictions
  with the experimental data} (e.g. those discussed in~\cite{Harko}),
  because ordinary $M^{BEC}\equiv M^{(0)}$ leads to some overestimation.

 Note that $M^{(\mu)}$, the result of deformation,
remains the temperature dependent function while $R$ in (45) does
not. This is related with re-definition of the particle density and
the (critical) volume per particle in the case of deformed
thermodynamics: $v_c=\lambda^3/g^{(\mu)}_{3/2}(1)$.

\section{Concluding remarks}

 Main thermodynamic quantities of $\mu$-Bose gas -- total mean number
 of particles $N^{(\mu)}$, the ($\log$ of) the $\mu$-deformed partition function etc.,
 involve the $\mu$-generalizations $g^{(\mu)}_{k}(z)$ of polylogarithms
$g_{k}(z)$. This fact has influence on other results in this paper.
The metric tensor, Christoffel symbols and hence the Riemann
curvature get expressed through $\mu$-polylogarithms. Positive sign
of curvature and its divergence as $z\!\to\!1$ witness the
attraction between particles and the evidence of Bose-like
condensation. Formula for the $\mu$-dependent critical temperature
$T_c^{(\mu)}$ is compared with usual Bose case, with infinite
statistics system, and with $T_c^{(p,q)}$ of $p,q$-Bose gas model
(in the latter, $T_c^{(\mu)}$ can be both higher and lower,
depending on the values of parameters $p$ and $q$).
 The ratio $T_c^{(\mu)}/{T_c}$ as a function of $\mu$-parameter
 shows: \ critical temperature $T_c^{(\mu)}$ is higher than critical
$T_c$ of usual Bose gas, in contrast with the system of infinite
statistics which possesses~\cite{Infinite} lower critical
temperature than the $T_c$ of usual Bose gas. However, such a
property seems not to be a drawback, from the viewpoint of possible
application of $\mu$-Bose gas as modeling dark matter: in fact, in
our case stability of the condensate extends even higher in
temperature than usual $T_c$. Also, it can be shown that other facts
important for the modeling and confirmed for the case of infinite
statistics~\cite{Infinite} -- e.g. the smallness of particle mass,
can be examined for the $\mu$-bosons as well.

In the context of dark matter, the inner structure of its
constituents at a given deformation plays a remarkable role in their
response to extrinsic perturbation.
     The parameter $\mu$ is determined by specific conditions of the dark matter
existence, in each galaxy or a local region of Universe.
Moreover, since deformed critical temperature obeys the condition
$T^{(\mu)}_c\geq T^{(0)}_c>T$, we hope that the dark matter model
remains valid in interstellar environment where the temperature $T$,
concentration of the dark matter particles $\varrho$
($T_c\sim\varrho^{2/3}$), and parameter $\mu$ can vary.

 Other remarkable properties of $\mu$-Bose gas model (e.g. the falling behavior of
 entropy-per-volume versus the $\mu$-parameter, that means decreasing
 chaoticity with growing deformation strength), can also lead to interesting
 implications concerning the use for modeling dark matter.

   We conclude that, like infinite statistics particles, the $\mu$-Bose gas
 model has its own virtues and thus can be used to effectively model basic
 properties of dark matter.
   In that domain, the model may turn out to be just as successful as
($\tilde{\mu},q$)-deformed analog of Bose gas model in the effective
description, see Fig.~5 in~\cite{GM_NP}, of the unusual non-Bose
like properties of (the intercepts of) two-pion correlations
observed in the STAR/RHIC experiments.
   Obviously, further steps and more detailed study are needed in order
 to put the proposal on firm ground.


\section{Appendix: Infinite Statistics}

Gas of particles obeying~\cite{Greenberg,Medvedev} infinite
statistics (with parameter $p$) was proposed in \cite{Infinite} as a
model of dark matter.
   Authors considered thermodynamical geometry of this gas,
 and showed with their fig.~1 that condensation does occur
 for such system\footnote{Note that the figure in \cite{Infinite}
 was probably obtained numerically, as there are no explicit expressions
 of calculated metric components, Christoffel symbols and curvature
 in that paper. For that reason and for the sake of comparison with
 the results obtained in our $\mu$-Bose gas model, we present here
 explicitly the needed geometric quantities, all being expressed in terms of
 Lerch transcendent~\cite{Lerch}.}.

Number of particles and internal energy for infinite statistics gas
are given \cite{Infinite} as
\begin{equation}\label{energy}
U = \frac{A}{\beta^{\frac{d}{2}+1}}\int\limits_0^{\infty}
\frac{4pze^xx^{\frac{d}{2}}}{e^{2x}-p^2z^2}dx \ ,
\end{equation}
\begin{equation}\label{numpart}
N = \frac{A}{\beta^{\frac{d}{2}+1}}\int\limits_0^{\infty}
\frac{4pze^xx^{\frac{d}{2}-1}}{e^{2x}-p^2z^2}dx \ .
\end{equation}
 One can show that these functions can be expressed through
 the Lerch transcendent~\cite{Lerch} whose definition is
\begin{equation}
\Phi(z,s,\alpha) = \frac{1}{\Gamma(s)}
\int\limits_0^{\infty}\frac{t^{s-1}e^{-\alpha t}}{1-ze^{-t}} dt
\end{equation}
(here and below $\Gamma(x)$ is the usual gamma-function).
  The expressions (\ref{energy}) and (\ref{numpart}) are rewritten as
\begin{equation}
U = \frac{A\,\Gamma(\frac{d}{2}+1)}{(2\beta)^{\frac{d}{2}+1}}4pz
\Phi\left(p^2z^2, \frac{d}{2}+1, \frac{1}{2}\right) \ ,
\end{equation}
\begin{equation}
N = \frac{A\Gamma(\frac{d}{2})}{(2\beta)^{\frac{d}{2}+1}}4pz
\Phi\left(p^2z^2, \frac{d}{2}, \frac{1}{2}\right) \ .
\end{equation}
Using the known identity \cite{Lerch} for Lerch transcendent
  $$ 
\Phi(z,s-1,\alpha) = \left(\alpha + z\frac{d}{dz}\right)
\Phi(z,s,\alpha) \ ,
  $$ 
we calculate the metric components in the space of two parameters
$(\beta,\gamma)$ in Fisher-Rao representation
  $$  
G_{\beta\beta} =
\frac{A\,\Gamma\left(\frac{d}{2}+1\right)}{(2\beta)^{\frac{d}{2}+1}}\,8pz\,
\Phi\left(p^2z^2, \frac{d}{2}+1, \frac{1}{2}\right) \ ,
  $$  
  $$  
G_{\beta\gamma} =
\frac{A\,\Gamma(\frac{d}{2}+1)}{(2\beta)^{\frac{d}{2}+1}}\,8pz
\,\Phi\left(p^2z^2, \frac{d}{2}, \frac{1}{2}\right)\ ,
\hspace{0.7cm}
  $$  
  $$  
G_{\gamma\gamma} =
\frac{A\,\Gamma(\frac{d}{2})}{(2\beta)^{\frac{d}{2}}}
\,8pz\,\Phi\left(p^2z^2, \frac{d}{2} -1, \frac{1}{2} \right)
\hspace{0.8cm} ,
  $$  
 and the metric determinant:
$$
\det|G_{ij}| =
\frac{A^2(8\xi)^2}{(2\beta)^{d+2}}\left(\Gamma\left(\frac{d}{2}+2\right)\Gamma
   \left(\frac{d}{2}\right)\Phi_{-1}\Phi_1 - \right.
$$
  $$  
  \hspace{-15mm}
\left. -\Gamma^2\left(\frac{d}{2}+1\right) \Phi_0^2 \right) \ .
  $$  
Denote $pz \equiv \xi$. The Christoffel symbols are found to be
  $$  
\Gamma_{\beta\beta\beta} = -  \frac{A}{(2\beta)^{\frac{d}{2}+3}}
\Gamma \left(\frac{d}{2}+3\right) 8\xi \,\Phi
\left(\xi^2,\frac{d}{2}+1,\frac{1}{2}\right) ,
  $$  
  $$  
\Gamma_{\beta\beta\gamma} = -  \frac{A}{(2\beta)^{\frac{d}{2}+2}}
\Gamma
   \left(\frac{d}{2}+2\right) 8\xi \,\Phi
   \left(\xi^2,\frac{d}{2},\frac{1}{2}\right),
   \hspace{6mm}
  $$  
  $$  
\Gamma_{\beta\gamma\gamma} = -  \frac{A}{(2\beta)^{\frac{d}{2}+1}}
\Gamma \left(\frac{d}{2}+1\right) 8\xi \,\Phi
 \left(\xi^2,\frac{d}{2}-1,\frac{1}{2}\right) ,
  $$  
  $$  
\Gamma_{\gamma\beta\beta} = -  \frac{A}{(2\beta)^{\frac{d}{2}+2}}
\Gamma \left(\frac{d}{2}+2\right) 8\xi \,\Phi
 \left(\xi^2,\frac{d}{2},\frac{1}{2}\right) ,
   \hspace{6mm}
  $$  
  $$  
\Gamma_{\gamma\beta\gamma} = -  \frac{A}{(2\beta)^{\frac{d}{2}+1}}
\Gamma \left(\frac{d}{2}+1\right) 8\xi\,\Phi
 \left(\xi^2,\frac{d}{2}-1,\frac{1}{2}\right) ,
  $$  
  $$  
\Gamma_{\gamma\gamma\gamma} = -  \frac{A}{(2\beta)^{\frac{d}{2}}}
\Gamma \left(\frac{d}{2}\right) 8\xi \,\Phi
 \left(\xi^2,\frac{d}{2}-2,\frac{1}{2}\right) .
   \hspace{9mm}
  $$  
Using the notation
  $$  
\Phi \left(\xi^2,\frac{d}{2}+k,\frac{1}{2}\right) \equiv \Phi_k
  $$  
the resulting expression for Riemann tensor is given as
$$
R_{\beta\gamma\beta\gamma} = \frac{A}{ (2\beta)^{\frac{d}{2}+2}}
\Gamma \left(\frac{d}{2}+2\right) \Gamma \left(\frac{d}{2}+1\right)
   \Gamma \left(\frac{d}{2}\right) 8\xi \times \hspace{0.5cm}
$$
  $$  
  \hspace{-14mm}
\hspace{1.5cm} \times \frac{\left(-2 \Phi_1 \Phi_{-1}^2+\Phi_0^2
\Phi_{-1}+\Phi_{-2} \Phi_1 \Phi_0 \right)}{\Gamma
\left(\frac{d}{2}+2\right) \Gamma
   \left(\frac{d}{2}\right) \Phi_{-1}  \Phi_1 -\Gamma \left(\frac{d}{2}+1\right)^2
   \Phi_0^2}
  $$  
and the thermodynamical curvature as
$$
R = \frac{(2\beta)^{\frac{d}{2}+2}}{ 4\xi A} \Gamma
\left(\frac{d}{2}+2\right) \Gamma \left(\frac{d}{2}+1\right)
   \Gamma \left(\frac{d}{2}\right) \times \hspace{1cm}
$$
\begin{equation}
\hspace{8mm} \times \frac{ \left(-2 \Phi_1 \Phi_{-1}^2+\Phi_0^2
\Phi_{-1}+\Phi_{-2} \Phi_1 \Phi_0 \right)}{\left(\Gamma
\left(\frac{d}{2}+2\right) \Gamma
   \left(\frac{d}{2}\right) \Phi_{-1}  \Phi_1 -\Gamma \left(\frac{d}{2}+1\right)^2
   \Phi_0^2\right)^2} .
\end{equation}
As it is seen from fig.~1 in \cite{Infinite} and also follows from
eq.~(51), in three dimensional ($d\!=\!3$) space sending $ zp\to 1 $
results in $R \to \infty $ and thus leads to the phase transition
(Bose-like condensation). This fact, and some other properties of
particles with infinite statistics (their mass, weakness of their
interaction etc.) has allowed the authors of \cite{Infinite} to
conclude in favor of ability of such system as possible model of
dark matter.

\end{document}